\documentstyle[preprint,aps,psfig]{revtex}
\tighten
\preprint{\vbox{ \hfill IU/NTC
    97-12 \\
   \null \hfill}}
\begin{document}

\draft
\title{Parity Violating Elastic Electron Scattering and
Coulomb Distortions}
\author{C. J. Horowitz\footnote{email: charlie@iucf.indiana.edu}}
\address{Nuclear Theory Center \\
2401 Milo B. Sampson Lane \\
Bloomington, Indiana 47405 USA}

\date{\today}

\maketitle

\begin{abstract}
Parity violating elastic electron-nucleus  scattering provides an accurate 
and model independent measurement of neutron densities, because 
the $Z^0$ couples primarily to neutrons.  Coulomb distortion 
corrections to the parity violating asymmetry $A_l$ are calculated 
exactly using a relativistic optical model.  Distortions significantly 
reduce $A_l$ in a heavy nucleus.  However even with distortions, an 
experiment to measure the neutron radius is feasible.  
This will aid the interpretation of 
future atomic parity violation measurements and provide fundamental
nuclear structure information.  Coulomb distortions and small differences
between neutron and proton radii could be important for a standard model
test on $^4$He, $^{12}$C or $^{16}$O.
\end{abstract}

\bigskip
\section{Introduction}

Parity violating electron-nucleus scattering is important for 
several reasons.  First, it provides a test of the standard model at low 
energies. Indeed, an early experiment on $^{12}$C was performed [1]
(with somewhat limited accuracy).  Second, it is sensitive to strange quarks 
in the nucleon or nucleus.  Proposed Jefferson Laboratory experiments hope
to extract strange quark contributions to the electric form factor of the 
nucleon from elastic scattering on $^4$He [2].  Finally, parity violation
provides a unique and very clean way to study neutron densities and isospin
violation in nuclei [3].   This is because the $Z^0$ couples predominantly to
neutrons.  [Note, the $Z^0$ proton coupling depends on the small factor 
$1- 4{\rm sin}^2\Theta_W$.]

An accurate measurement of neutron distributions in a 
heavy nucleus would provide fundamental nuclear structure 
information.  It will constrain isovector terms in the nuclear matter energy
functional such as the surface symmetry energy.  This could be 
important in astrophysics when one extrapolates to unstable very 
asymmetric nuclei.  Furthermore, a measurement
of neutron radii will significantly aid the interpretation of future 
atomic parity violation measurements.  The present $Cs$ experiment is 
accurate to 0.3 percent [4].  Pushing the accuracy of atomic experiments 
is important as a test of the standard model and as a search for new 
physics.  
However, a 0.1 percent measurement in a heavy atom will require knowing the 
neutron radius to of order one percent [5]  (to keep
uncertainties in the neutron density from interfering with a 
standard model test).

The charge density is known from elastic electron scattering.  
Thus, to determine the neutron radius to one percent requires 
knowing the difference between neutron and proton radii 
to about 25 percent.  This accuracy is probably beyond that of present 
nuclear theory (we comment on this below).  Furthermore, neutron radii 
determinations from hadronic probes suffer from large systematic errors.
Therefore, a measurement of the parity violating asymmetry for elastic 
electron scattering should provide crucial information for
atomic parity experiments.  

Atomic experiments also depend on atomic theory (the overlap of electronic
wave function with the nucleus) which at present is only good to about
one percent in $Cs$ [6].  However, this accuracy may improve in the future.
If this is not the case, atomic experiments may shift to measuring ratios of
parity violation in different isotopes because many of the atomic 
uncertainties cancel.  Isotope ratios place much more stringent requirements
on knowledge of the neutron radius and how this changes among isotopes.
This knowledge is beyond present nuclear theory.  However, an accurate 
measurement of the neutron radius with electron scattering on a single 
nucleus should still provide an important {\it first step} towards 
calibrating a theory of neutron radii differences. 

Electron scattering from a heavy nucleus is modified substantially by coulomb
distortions.  These effects are of order $Z\alpha$ (where $Z$ is the nuclear 
charge) and will modify the parity violating asymmetry.  However, there are 
no previously published calculations.  In the
present paper we accurately calculate coulomb distortion effects with a 
relativistic optical model. The Dirac equation is numerically solved for
an electron moving in vector and axial vector potentials.  Our formalism
is presented in section II along with checks of the numerics. 

Elastic parity violating asymmetries from several nuclei are shown in 
section III.  Results are also shown for a variety of electron energies.  We 
conclude in section IV that coulomb distortions significantly modify the 
asymmetry.  However these are accurately calculated.  Even with
distortions, the asymmetry is very sensitive to neutron densities and
an experiment to measure the neutron radius in a heavy nucleus is 
feasible.  We also conclude that coulomb distortions are important for a one 
percent standard model test in $^{12}C$ or $^{16}O$ and that such a test may 
be sensitive to very small differences in proton and neutron radii. 
\bigskip
\section{Formalism}
\medskip
In this section we describe our relativistic optical model formalism and 
discuss a variety of checks on our numerical results.  
The electron wave function $\Psi$ (for scattering from a spin zero nucleus) 
is assumed to satisfy a Dirac equation,
$$[{\bf \alpha}\cdot {\bf p} + \beta m_e + \hat V(r)]\Psi = E \Psi.\eqno(1)$$
Here $E$ is the center of mass energy and we neglect other center of mass 
corrections.  The total potential $\hat V(r)$ has vector $V(r)$ and axial 
vector $A(r)$\ components,
$$\hat V(r) = V(r) + \gamma_5 A(r).\eqno(2)$$
The conventional coulomb potential is $V$\ while weak neutral currents give 
rise to $A$ which is of order the Fermi constant $G_F$,
$$A(r)= {G_F\over 2^{3/2}} \rho_W(r).\eqno(3)$$
The weak charge density $\rho_W$ is closely related to (minus) the neutron 
density (see below) and is normalized, for neutron number $N$ and proton 
number $Z$,
$$\int d^3r \rho_W(r)= -N + (1-4{\rm sin}^2\Theta_W) Z.\eqno(4)$$
Equation (1) includes terms of all orders in $Z\alpha$.  This is important 
because $Z\alpha$\ is large for a heavy nucleus.  Equation (1) neglects 
radiative corrections, which are higher order in $\alpha$, and dispersion 
corrections where the intermediate nucleus is in an excited state.

In the limit of vanishing electron mass, it is a simple matter to include the 
effects of the axial potential $A(r)$.  One writes the Dirac equation 
for helicity states with $\Psi_\pm={1\over 2}(1\pm\gamma_5)\Psi$,
$$[{\bf \alpha}\cdot {\bf p} + V_\pm(r)]\Psi_\pm = E \Psi_\pm,\eqno(5)$$
and
$$V_\pm(r) = V(r) \pm A(r).\eqno(6)$$ 
Thus, the positive helicity state scatters from a potential $V+A$ while 
the negative helicity state scatters from $V-A$.  To calculate the parity 
violating asymmetry $A_l$ one simply calculates the scattering amplitudes 
for $V+A$ and $V-A$ and subtracts,
$$A_l={d\sigma_+/d\Omega - d\sigma_-/d\Omega \over
       d\sigma_+/d\Omega + d\sigma_-/d\Omega}.\eqno(7)$$

We have written a new relativistic optical code ELASTIC which numerically 
solves the partial wave Dirac equation and sums up phase shifts to 
calculate the scattering amplitude.  From the amplitude it calculates the 
following observables: the unpolarized cross section, the parity conserving 
analyzing power $A_y$ (for an initial electron spin normal to the reaction 
plane), the parity violating asymmetry $A_l$ and the spin rotation parameter 
$Q$.  This is related to the angle through which the electron's spin is 
rotated when it scatters from the nucleus [7].  The 
normal analyzing power $A_y$ is of order $m_e/E$ and vanishes in Born approximation.
It is very small, comparable to $A_l$!  We will discuss $A_y$ in a future paper.

The numerical details of the code will also be presented in a later paper.  
Here we describe some of the checks which give us confidence in our results.  
First the code must reproduce known cross sections.  For example, 
elastic cross sections from $^{208}Pb$ at 502 MeV are reproduced out to 3.7 
Fm$^{-1}$ (the extent of the data [8]).  Near 3.7 Fm$^{-1}$ the error is of 
order a percent.  The cross section at these large angles is reduced by many 
orders of magnitude.  This requires that the scattering amplitude be 
calculated very accurately: indeed, more accurately then needed for the
forward angle $A_l$ (see below).

The code has been checked against plane wave results.  We 
multiplied both $V$ and $A$ in Eqs. (1,2) by a  small factor, say 0.01.  
Then the full code was run summing over many partial waves.  Finally the 
resulting cross section was divided by $0.01^2$ and both the cross section 
and $A_l$ were compared to know plane wave results. This check was performed 
both for $\rho_W$\ proportional to the charge density (where $A_l$ 
is linear in $q^2$) and for different neutron and proton densities.  The 
numerical agreement is very good (better then 0.1 percent for $A_l$) except 
right in the diffraction minima.  In the minima, the exact result should 
be different from plane wave results even for a system with the small 
charge of $0.01Z$.  This plane wave test is actually more demanding 
then the full calculation because some numerical errors are 
magnified in comparison to the small interaction. 

The spin rotation parameter $Q$ should be just,
$$Q={\rm sin}(\Theta),\eqno(8)$$
with $\Theta$ the electron scattering angle, up to small corrections 
of order $m_e/E$.  This is a nontrivial check since we must sum up 
over all partial waves to calculate $Q$.  Equation (8) is reproduced 
by our code except at very large momentum transfers
(beyond 4 $Fm^{-1}$ for $^{208}Pb$ where the cross section is 
also inaccurate).

Finally, the only new feature of the $A_l$ calculation is a subtraction of the
positive and negative helicity amplitudes.  In practice this is not a 
problem because the amplitude must be calculated to much better accuracy 
then a part in $10^5$ (a typical size for $A_l$) in order to reproduce 
the large angle cross section.  Furthermore, many errors cancel in the 
subtraction.  Nevertheless, this can be tested by multiplying just 
$A$ in Eq. (3) by $0.1$ (keeping $V$ unchanged) and running the full code.  
The resulting $A_l$ is scaled up by a factor of ten and seen to agree well 
with earlier results.  This procedure makes the subtraction ten times more 
sensitive and verifies its accuracy. 
Note, the code calculates observables to all orders in both 
$V$ and $A$.  In practice, $A$ is small so $A_l$ is linear in $A$.   
Therefore, the code can be run with almost any value of $G_F$ in 
Eq. (3) and the resulting output $A_l$ scaled appropriately. 

Taken together, these four tests check almost all areas of the calculation 
and give us confidence in our results.  In practice the calculation is 
no harder then older work for the unpolarized cross section.  Indeed, 
in a helicity basis only very small modifications 
are needed to include a parity violating potential.

\bigskip
\section{Results}
\medskip
In this section we present results of the code ELASTIC.  We first 
assume the weak density $\rho_W(r)$ has the same spatial distribution 
as the charge density $\rho(r)$.  This allows one to see the effects of 
only coulomb distortions.  Then we show results for different weak and 
electromagnetic densities.  To start, we use a simple three 
parameter Fermi charge density for $^{208}$Pb from ref. [9], see Table I.  
This fits all but the back angle electron scattering data.  The weak density 
is assumed to be proportional to the charge density,
$$\rho_W(r)= -[{N\over Z}+4{\rm sin}^2\Theta_W-1]\rho(r).\eqno(9)$$
This satisfies the normalization condition of Eq. (4).  For simplicity in
notation, we refer to the weak density given by Eq. (9) as being equal to
the charge density.

Figure 1 shows the asymmetry $A_l$ for $^{208}$Pb versus momentum 
transfer $q$ both in a plane wave impulse approximation where,
$$A_l=\bigl[ {G_F q^2\over 4\pi \alpha 2^{1/2}} \bigr] \bigl[{N\over Z} 
+ 4{\rm sin}^2 \Theta_W-1\bigr],\eqno(9b)$$
and then including full distortions at electron energies from 502 to 
3000 MeV.  Coulomb distortions are seen to reduce $A_l$ substantially, 
especially in the diffraction minima.   As the energy 
increases, the effects of coulomb distortions do not decrease (very much) 
instead there is a slight shift in the position of the diffraction
minima to higher momentum transfers.  We conclude from Fig. 1 that 
coulomb distortions must be included for parity violation in a heavy nucleus.

Figure 2 shows $A_l$ for $^{12}$C at 200 MeV.  This is the energy of 
the original BATES experiment[1].  This figure uses a relativistic mean 
field model [10](MFT) for the charge density.   For the MFT  we
approximate the weak density as,
$$\rho_W(r)=\int d^3r^\prime G_E(|{\bf r}-{\bf r}^\prime|)
[-\rho_n(r^\prime) + (1-4{\rm sin}^2\Theta_W)\rho_p(r^\prime)].\eqno(10)$$
Here $\rho_n$ and $\rho_p$ are point neutron and proton densities and the
electric form factor of the proton is approximated $G_E(r)\approx {\Lambda^3
\over 8\pi}e^{-\Lambda r}$ with $\Lambda=4.27$ Fm$^{-1}$.  This neglects
strange quark contributions, the neutron electric form factor and meson
exchange currents. It also assumes good isospin for the nucleon.  For 
simplicity, all calculations in this paper use sin$^2\Theta_W=0.23$ for the
Weinberg angle.

The dotted curve in Fig. 2 assumes 
Eq. (9) while the dashed curve uses the MFT weak density.  Both of
these are plane wave calculations.  Finally, the solid curve uses 
the MFT weak density and includes coulomb distortions.  In the MFT the 
protons have a slightly larger radius then the neutrons
because of coulomb repulsion.  This small change in radius 
can lead to a large change in $A_l$ at back angles.  At the 30 deg. 
angle of the BATES experiment the MFT plane wave 
calculation is about 1 \% above the equal density plane wave result.
Coulomb distortions increase $A_l$ by another two \%.  
Thus the full calculation is about 3 \% above the original prediction.
This change is smaller then the BATES error.  However, it 
is large compared to a possible one \% standard model test.

We conclude that coulomb distortions must be included in a one \% standard 
model test on $^{12}$C or $^{16}$O (indeed Fig. 3 shows similar 
results for $^{16}$O).  However, we 
have calculated coulomb distortions accurately so 
they should pose no problems for the interpretation of the experiment.  
We also see that isospin violation (small differences between proton 
and neutron densities) is significant
especially at back angles.  This correction involves some nuclear structure 
uncertainties.  Thus isospin violation
may limit a standard model test to small momentum transfers.

Figure 4 shows $A_l$\ for $^4$He at 850 MeV.  We assume a 3 parameter 
Fermi charge density, see Table I. For this light target, coulomb 
distortions are 
only important in the diffraction minima.  The solid curve is for a 
neutron density arbitrarily one \% smaller then the proton density.  
This change in $r_n$ is somewhat bigger then theoretical estimates. 
However, there is great sensitivity to small changes in the 
neutron density.  An accurate microscopic calculation of $r_n-r_p$ 
using Greens Function Monte Carlo or other methods is very important.

We note that the solid curve crosses the dashed curve just 
beyond 50 degrees.  This is near the
second maximum in the form factor and corresponds to the 
kinematics of a planned experiment [2].
At this momentum transfer, $q$, the derivative of the cross 
section with $q$ goes to zero which
reduces some systematic errors (such as those from helicity 
correlated changes in $q$).  One can think of the derivative 
as being with respect to the dimension-less quantity $qr$ with 
$r$ the nuclear radius.  Thus the derivative of the cross section 
with respect to $r$ also vanishes at the same point.  This implies 
that the cross section and asymmetry will be insensitive to small changes in 
$r$ or $r_n-r_p$.  This minimizes the sensitivity to isospin 
violation.  However the sensitivity is large at other
momentum transfers.

Results for $A_l$ in $^{208}$Pb at 850 MeV are shown in Fig. 5 at
forward angles and in Fig. 6 at backward angles.  
[Note, all of the curves in these and remaining figures include coulomb 
distortions.]  The dotted 
curve assumes a three parameter Fermi charge density and equal weak density.  
The solid curve uses relativistic mean field (MFT) [10]
charge and weak densities.  The large 
difference between these curves indicates a strong 
sensitivity to the neutron radius or $r_n-r_p$.  Finally, the dashed curve 
assumes the weak density is a scaled (stretched) version of the (three parameter
Fermi) charge density,
$$\rho_W(r)= -[{N\over Z} + 4{\rm sin}^2\Theta_W -1] \lambda^3 
\rho(\lambda r).\eqno(11)$$
The scale parameter $\lambda=.9502$ is chosen to reproduce the MFT $r_n-r_p$.
These various densities are shown in Fig. 7.  Root mean square radii are
collected in table II.
The good agreement between the dashed and solid curves in Fig. 5 indicates 
that a forward angle measurement is primarily sensitive to the neutron radius 
and not to shell structure in the density (see Fig. 7).  The scaled three parameter 
Fermi density is very different from the MFT at small $r$.

The nucleus $^{138}$Ba provides a meeting ground between nuclear
and atomic physics.  There is interest in an atomic parity violation experiment
on the Ba ion.  At the same time, $^{138}$Ba has a relatively simple nuclear
structure and a large gap of about 1.5 MeV to the first excited state.  
[Unfortunately Cs isotopes have relatively complicated nuclear structure and low
first excited states.]  Thus Ba may be a good place to measure both the neutron
radius (with electron scattering) and atomic parity violation.
Figure 8 shows MFT weak and charge densities for $^{138}$Ba and Fig. 9 
presents parity violating asymmetries.  The sensitivity to the neutron 
radius is large, comparable to Pb.

Alternatively, one may be able to accurately calibrate nuclear theory with a
measurement of $r_n-r_p$ in $^{208}$Pb and then use theory to interpolate to
other nuclei of interest in atomic physics.  In a latter paper we will
discuss both the absolute errors in nuclear theory and the relative
errors in going from one nucleus to another.  It should be possible
to achieve the needed one percent relative error.  Thus one may be able to
understand bulk neutron radii throughout the periodic table with only a single
measurement.

From a nuclear structure point of view alone, an obvious choice is $^{208}$Pb
since this is such a good doubly closed shell nucleus with a simple structure,
a high first excited state, a large cross section and a large neutron excess.
We also show in Figs. 10 and 11 predictions for $^{48}$Ca since this nucleus has
a large fractional neutron excess.  Again, there is a large sensitivity to the
neutron radius.  However, the cross section for $^{48}$Ca is smaller 
then for $^{208}$Pb.

We now discuss the optimal kinematics for an experiment on 
$^{208}$Pb.  The figure of merit,
$$F=A_l^2d\sigma/d\Omega,\eqno(12)$$
is shown in Fig. 12 for 850 MeV.  This is strongly forward peaked since the
cross section falls rapidly with angle.  Thus experiments may only be 
feasible at forward angles.  Figure 5 suggests one is sensitive to the neutron
density at scattering angles near 6 and 12 degrees in the Lab.  
This is made quantitative in Fig. 13 where we plot the logarithmic 
derivative of the asymmetry with respect to the scale factor of the neutron
density, $\lambda$\ of Eq. 11 (equivalently with respect to the neutron radius),
$${d{\rm Log} A_l\over d {\rm Log }\lambda} = {\lambda \over A_l} \bigl(
{d A_l\over d\lambda}\bigr).\eqno(13)$$
Note, this is evaluated at $\lambda=.9502$.  The logarithmic derivative 
peaks around 3.2 near 7 degrees and around 10 near 14 degrees.  A value
of 3.2 means that a 3.2 percent measurement of $A_l$\ could determine 
the neutron radius to one percent (if other uncertainties are small).  

The product of the figure of merit and the logarithmic derivative is
also shown in Fig. 12.  This is large where $A_l$ can be accurately measured 
and is sensitive to the neutron radius.  The first maximum in this
product (near 4 degrees) is about 20 times the second maximum 
(near 12 degrees). Figure 12 is for a fixed beam energy of 850 MeV.
Results can be approximately scaled to other energies, $E$, by multiplying the
cross section (at fixed momentum transfer) by ($E$/850 MeV)$^2$.  

For example, an experiment at a fixed laboratory angle of six degrees is
illustrated in Fig. 14.  Note, this is an approximate figure since it is
based on distortion calculations at 850 MeV scaled to other energies.  
However, it should provide a good first orientation.  A measurement at 
six degrees is possible in Hall A at Jefferson Lab. with a septum magnet.  
Figure 14 has local maxima near 730, 1720 and 2600 MeV.   The product of 
figure of merit times log. derivative is a factor of 3.8 (15) lower at 1720 
(2600) MeV then at 730 MeV.

A measurement near 730 MeV is insensitive to possible uncertainties
in the surface thickness (of the neutron density).  However, the surface 
thickness may be well known from theory where it is constrained by the 
surface energy.  A measurement near 1720 MeV is more sensitive to
the neutron radius (see Fig. 13) so it may be less sensitive to other 
corrections or errors.  Therefore it would be very useful to measure both
points.  However, most of the information on the neutron radius
can be extracted from a single measurement.  If pushed for time, it is
most important to make a single accurate measurement (then two less accurate
ones).

\section{Conclusions}
In this paper we have calculated parity violating asymmetries for
elastic electron scattering including coulomb distortions.  We solve
a relativistic optical model for electron scattering in vector
and axial-vector potentials.  A series of plane wave and cross section
checks give us confidence in the numerical results.

Our most important conclusion is that a parity violation experiment to
measure the neutron density in a heavy nucleus is feasible.  Possible targets 
include $^{208}$Pb because of its simple structure, good closed shells,
and large neutron excess or $^{138}$Ba  because of the overlap of atomic 
physics interest and a relatively simple nuclear structure.  It is
straight forward to optimize the kinematics of such an experiment.  
However, one must include the large effects of coulomb distortions.
One possibility for $^{208}$Pb is to measure around six degrees and an
energy near 750 and or 1700 MeV.  We would be happy to provide more
detailed calculations upon request.

Future atomic parity experiments will require accurate 
knowledge of the neutron radius.
In a later paper, we explore how a single good electron scattering 
measurement, coupled with the
many constraints of nuclear theory, should be enough to predict the
neutron radius to one percent for all closed shell nuclei.  Note, determining
small differences between isotopes is clearly more demanding.  However,
an understanding of bulk neutron radii is still an important first step
towards a theory of isotope differences.  

A measurement of the neutron radius will also provide fundamental nuclear
structure information.  It would be the first accurate and model independent 
measurement of the {\it size} of large hadronic systems.  Note, the size 
does not follow directly from the charge radius because of the neutron skin.  
The measurement will provide important constraints on the isospin 
dependence of the nuclear matter energy functional and should constrain 
parameters such as the surface symmetry energy and or the isovector 
incompressibility.

We have found that coulomb distortions are also important for a one 
percent standard model test in $^{12}$C or $^{16}$O.  However, we have
calculated distortions accurately so they should not pose a problem in the 
interpretation of an experiment.  Small differences
between proton and neutron radii are also important for elastic experiments
involving $^{4}$He, $^{12}$C and $^{16}$O.  Microscopic calculations
of the difference between neutron and proton radii in $^{4}$He or $^{16}$O 
would be very useful.  

\section*{Acknowledgments}

I thank Tim Cooper for a great deal of very valuable help on the numerics.
Bill Donnelly, Steve Pollock and Mike Ramsey-Musolf are thanked for
useful physics discussions.  I thank Bunny Clark for electron scattering 
data and Dick Furnstahl and Horst Muller for neutron densities.
Supported in part by DOE grant number DE-FG02-87ER-40365.

\begin{table}
\caption{ Three parameter Fermi Densities [9] $\rho=\rho_o [1+ w (r/R)^2] /
 [1+ \exp((r-R)/a)] $ }
\begin{tabular}{cccc}
Nucleus    & R          & a       & W     \\
\tableline
& $Fm$      &$Fm$    & \\
$^4$He   & 1.008 & 0.327 &   0.445 \\
$^{208}$Pb & 6.4        & 0.54    & 0.32  
\end{tabular}
\end{table}

\begin{table}
\caption{Root mean square radii for densities used}
\begin{tabular}{cccc}
Nucleus & Density & Charge (Fm) & Weak (Fm) \\ \tableline
$^4He$     & 3p\footnote{Three parameter Fermi function, see Table 1. MFT = relativistic mean field theory densities from reference [10].}      & 1.717  & 1.717 \\
$^{12}$C   & MFT     & 2.504  & 2.477 \\
$^{16}$O   & MFT     & 2.753  & 2.720 \\
$^{48}$Ca  & MFT\footnote{Densities for $^{48}$Ca include a small correction from a nonzero neutron electric form factor $G^n_E$.} & 3.419  & 3.667 \\
$^{138}$Ba & MFT     & 4.797  & 5.038 \\
$^{208}$Pb  & 3p      & 5.490  & 5.490 \\
                      & 3p $(\lambda = .9502)$ & & 5.778 \\
           & MFT     & 5.456 & 5.744 \\
\end{tabular}
\end{table}

\begin{figure}
\centering{\ \psfig{figure=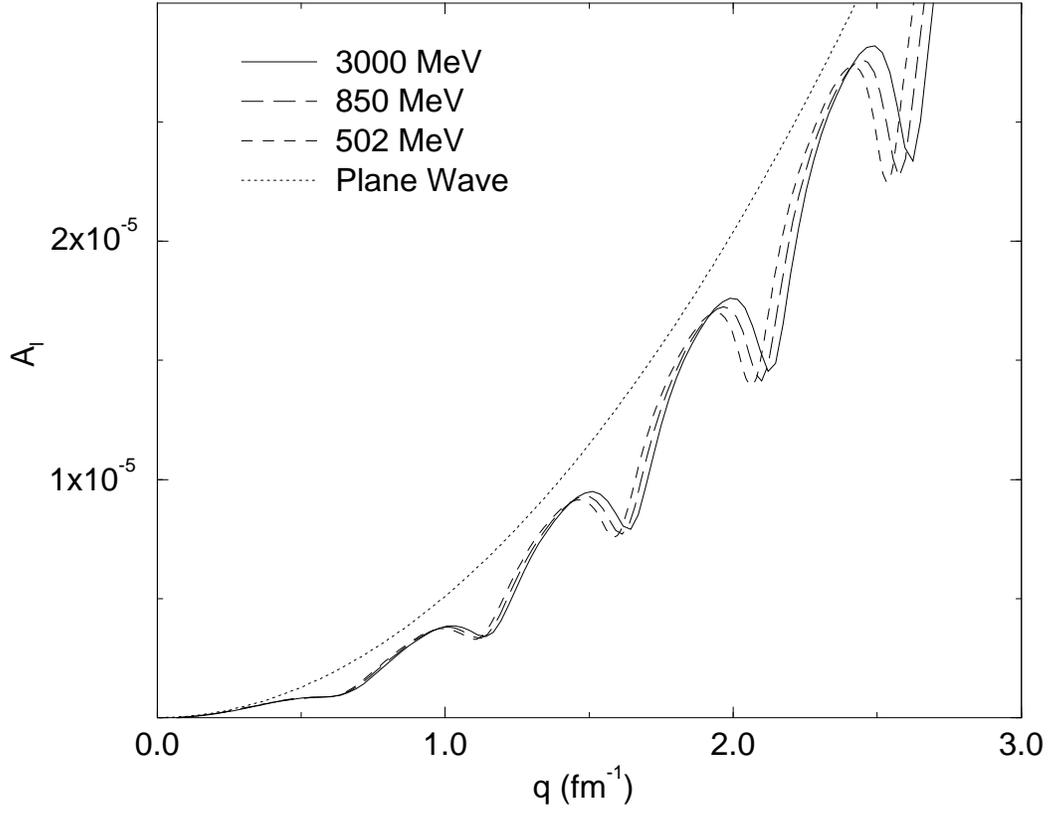,width=6in} }
\caption{Parity violating asymmetry $A_l$ for elastic 
scattering from $^{208}$Pb
vs. momentum transfer $q$ assuming
``equal" weak and charge densities (which are taken to be three
parameter Fermi functions) see Eq. (9).  The dotted curve is a plane wave 
approximation while full distorted wave results at 502 MeV are short dashed, 
850 MeV long dashed and 3000 MeV solid curves.}
\end{figure}

\begin{figure}[h]
\centering{\ \psfig{figure=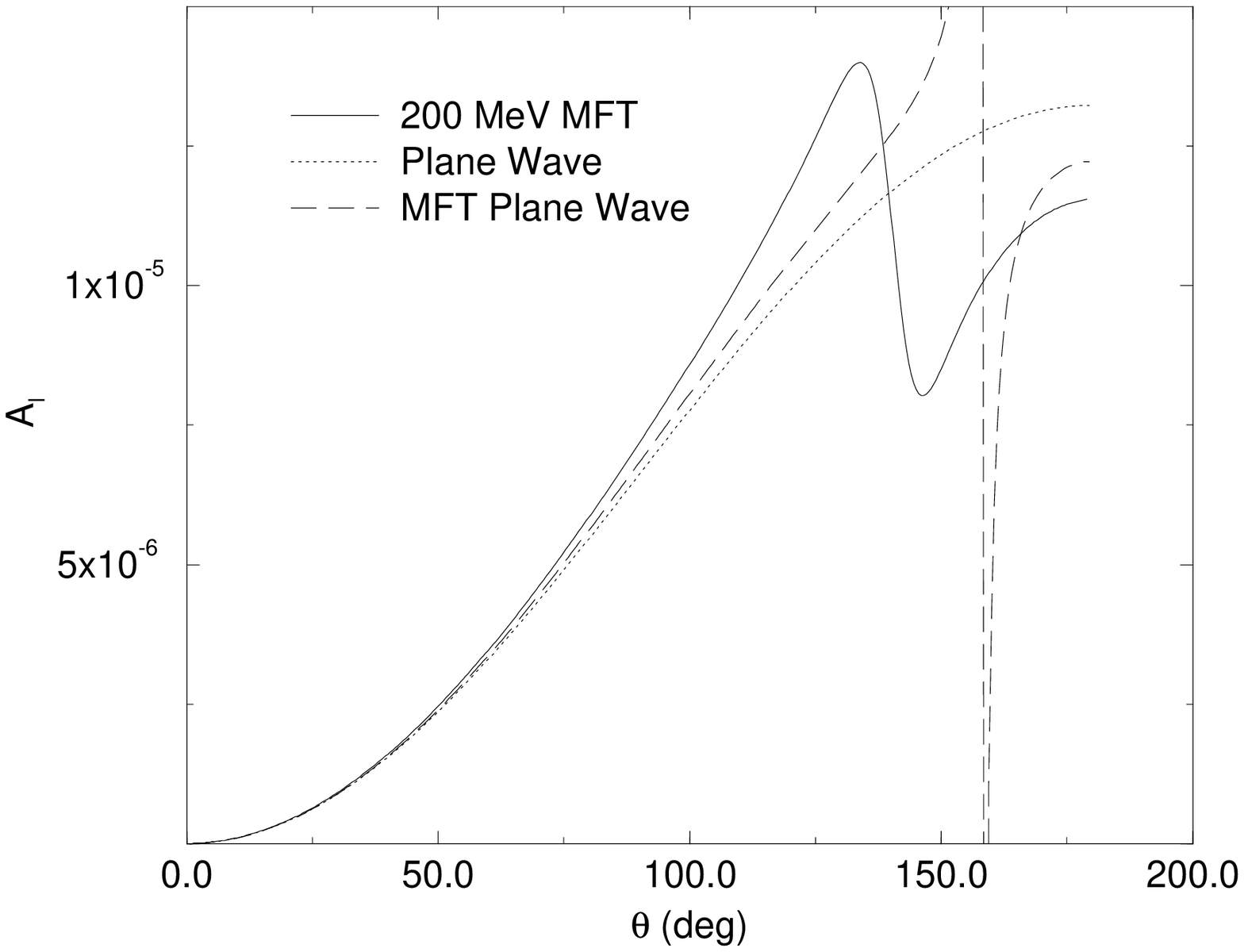,width=6in} }\nobreak
\caption{Parity violating asymmetry $A_l$ for elastic 
scattering from $^{12}$C
at 200 MeV vs. scattering angle $\theta$.  
Plane wave results using relativistic
mean field densities (which are slightly different for neutrons and protons) 
are the dashed curve while the dotted curve is a plane wave calculation 
assuming equal neutron and proton densities.  Finally, the solid curve 
is a full distorted wave calculation based on rel. mean field densities.}
\end{figure}

\begin{figure}[h]
\centering{\ \psfig{figure=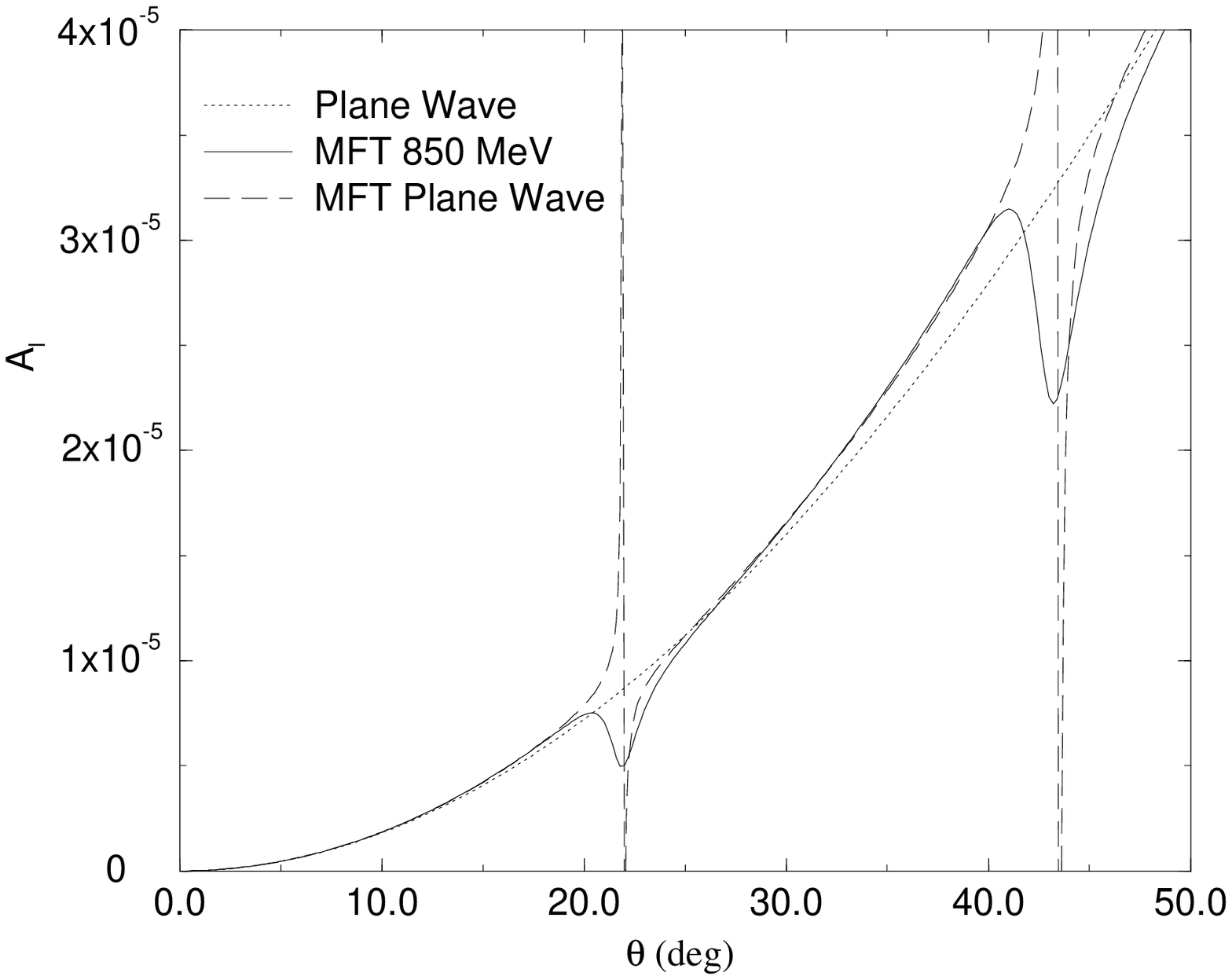,width=6in} }\nobreak
\caption{Parity violating asymmetry $A_l$ for elastic scattering 
from $^{16}$O at 850 MeV vs. scattering angle $\theta$.  
Plane wave results using relativistic
mean field densities (which are slightly different for neutrons and protons) 
are the dashed curve while the dotted curve is a plane wave calculation 
assuming equal neutron and proton densities.  Finally, the solid curve is 
a full distorted wave calculation based on rel. mean field densities.}
\end{figure}

\begin{figure}[h]
\centering{\ \psfig{figure=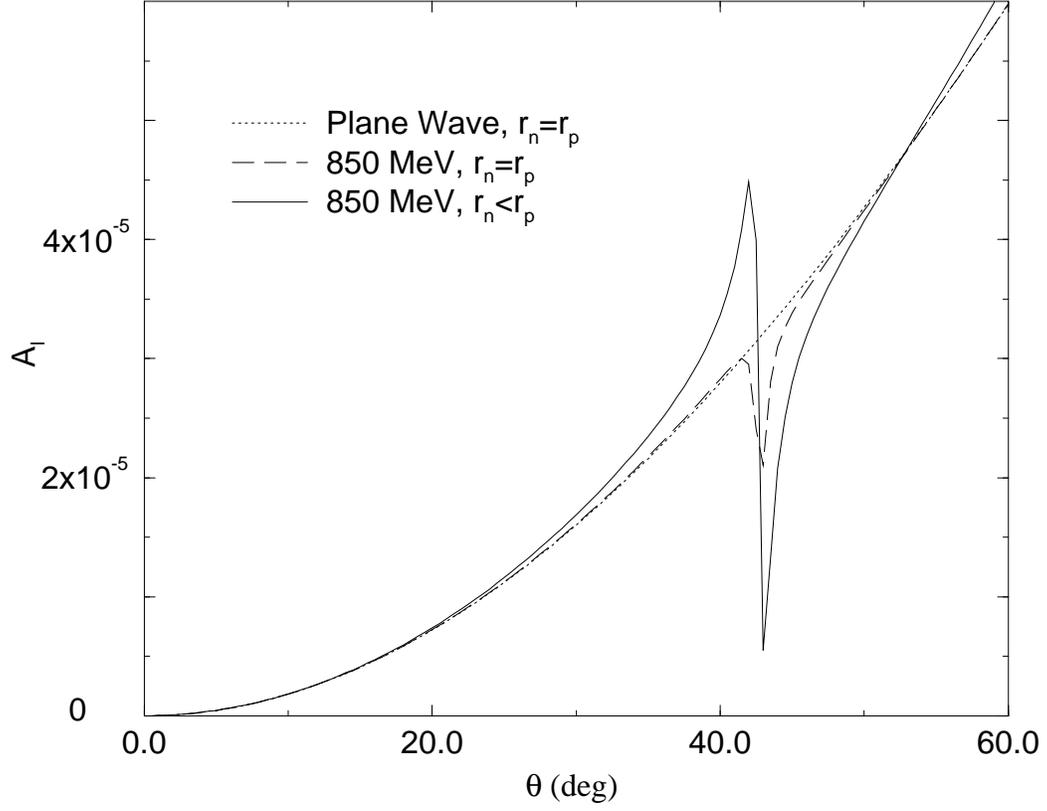,width=6in} }
\caption{Parity violating asymmetry $A_l$ for elastic scattering 
from $^{4}$He at 850 MeV vs. scattering angle $\theta$.  
Plane wave results using equal weak 
and charge densities (assumed to be a three parameter Fermi 
function) are the 
dotted curve.  Distorted wave calculations with equal weak and charge 
densities are dashed and the solid curve includes distortions assuming the
proton radius is one percent larger then the neutron radius.}
\end{figure}

\begin{figure}[h]
\centering{\ \psfig{figure=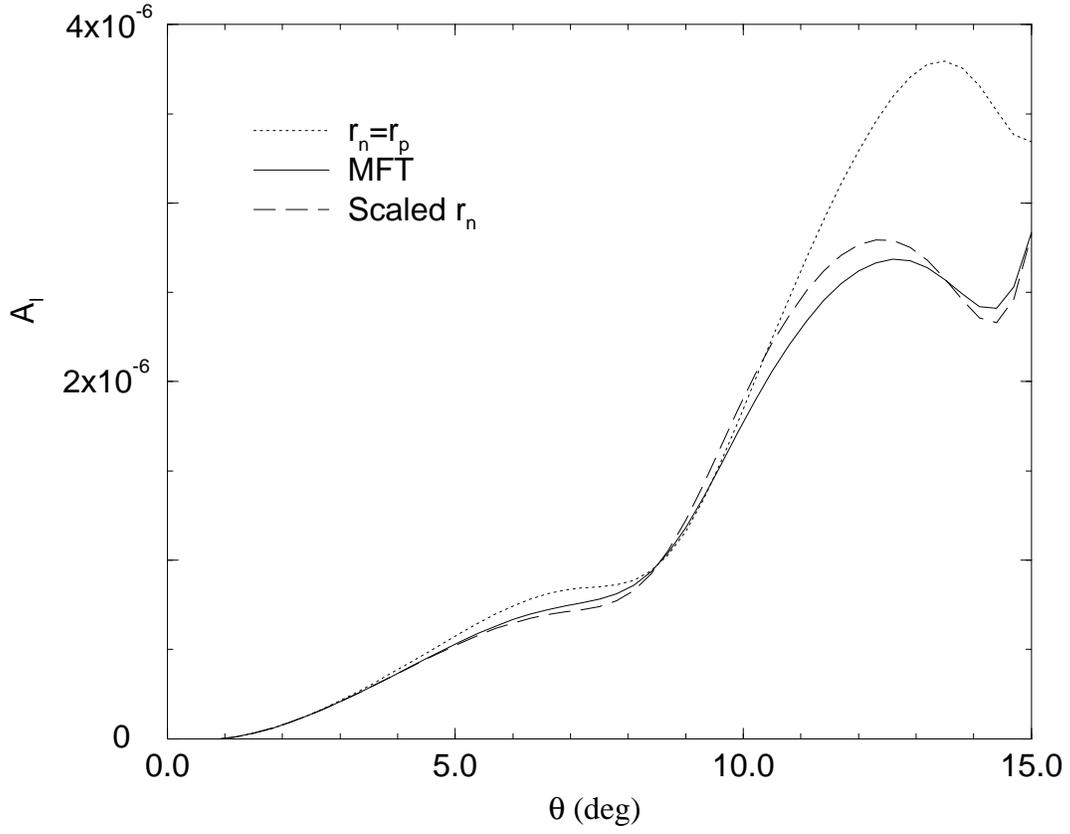,width=6in}}
\caption{Parity violating asymmetry $A_l$ for $^{208}$Pb at 850 
MeV vs. scattering angle $\theta$.  
The dotted curve uses equal weak and charge densities (assumed
to be three parameter Fermi functions) while the solid curve is based on 
relativistic mean field densities.  Finally the dashed curve assumes three
parameter Fermi densities, however the weak density has been stretched (with 
$\lambda=.9502$ see Eq. 11) to give the same difference in radii $r_n-r_p$
as the rel. mean field densities.  These densities are shown in Fig. 7.
Note, all curves in this and latter figures include distortions.}\end{figure}

\begin{figure}[h]
\centering{\ \psfig{figure=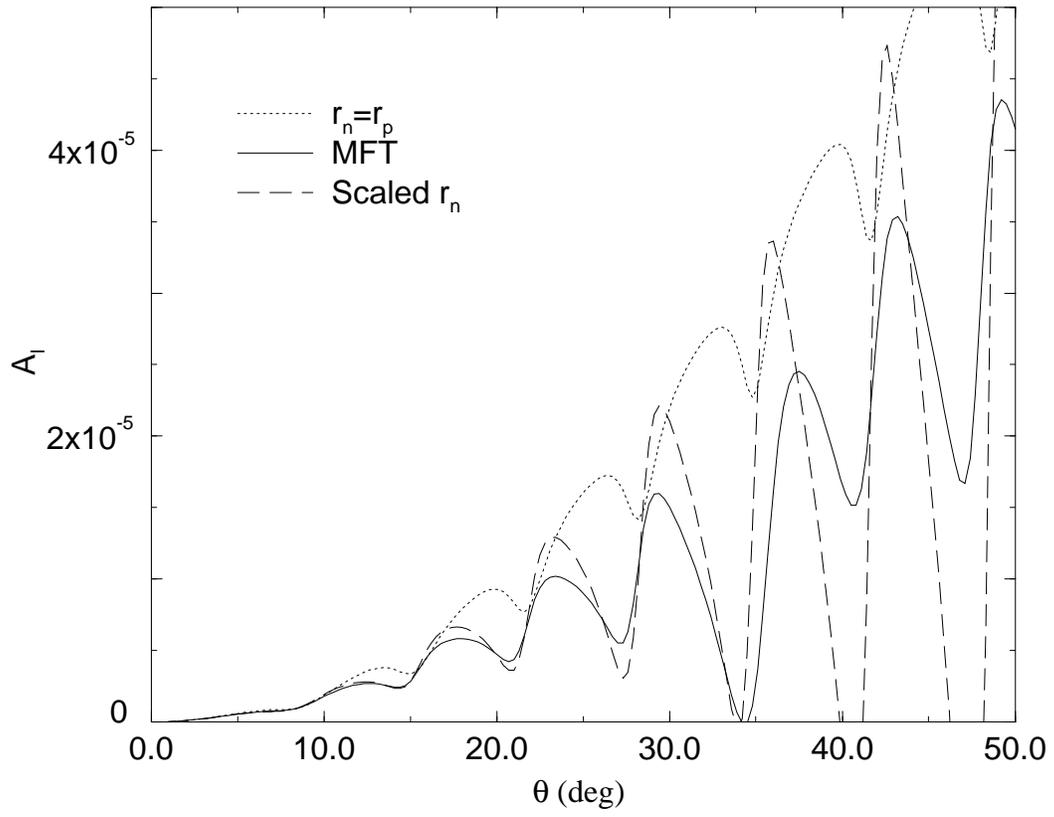,width=6in} }
\caption{As Fig. 5 except for larger scattering angles.}
\end{figure}

\begin{figure}[h]
\centering{\ \psfig{figure=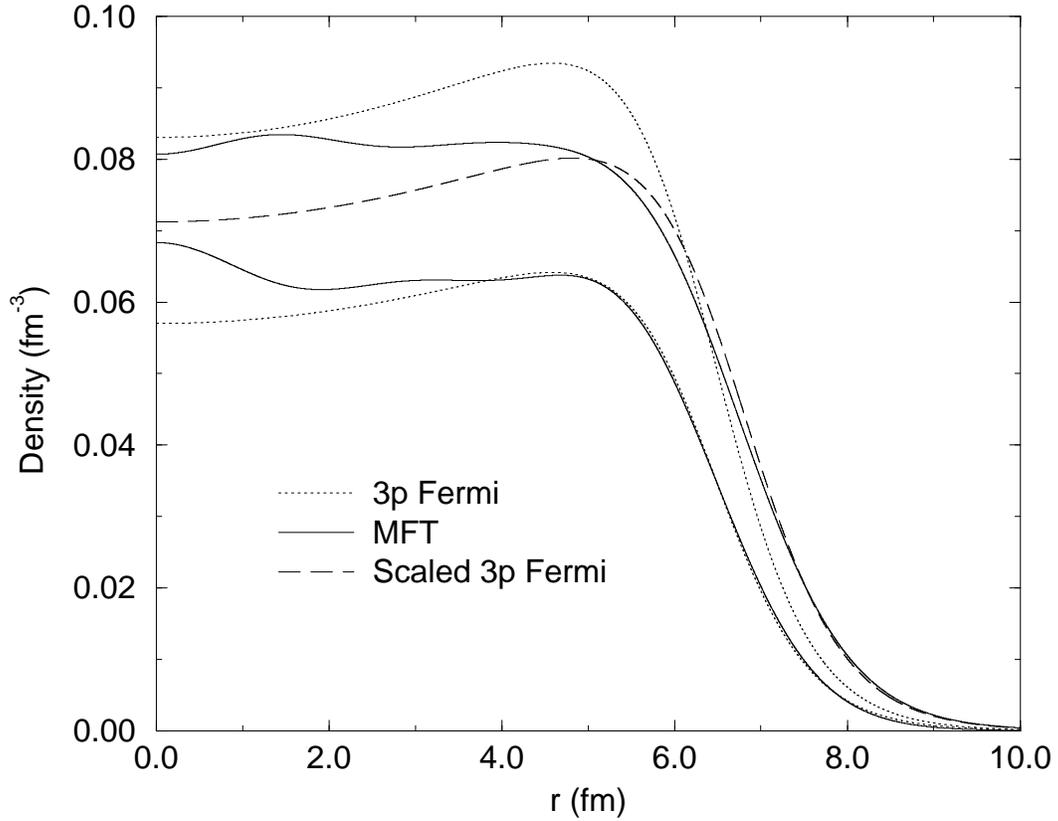,width=6in} }
\caption{Densities of $^{208}$Pb vs. radius $r$.  The lower two curves are
charge densities: the solid curve is the relativistic mean field result [10] 
while the dotted curve is a three parameter Fermi fit to elastic scattering.
The upper three curves are (minus the) weak density: solid, rel. mean field,
dotted, three parameter Fermi charge density normalized as in Eq. (9) while 
the dashed curve is this three parameter Fermi stretched by $\lambda=.9502$, 
see Eq. (11).}\end{figure}

\begin{figure}[h]
\centering{\ \psfig{figure=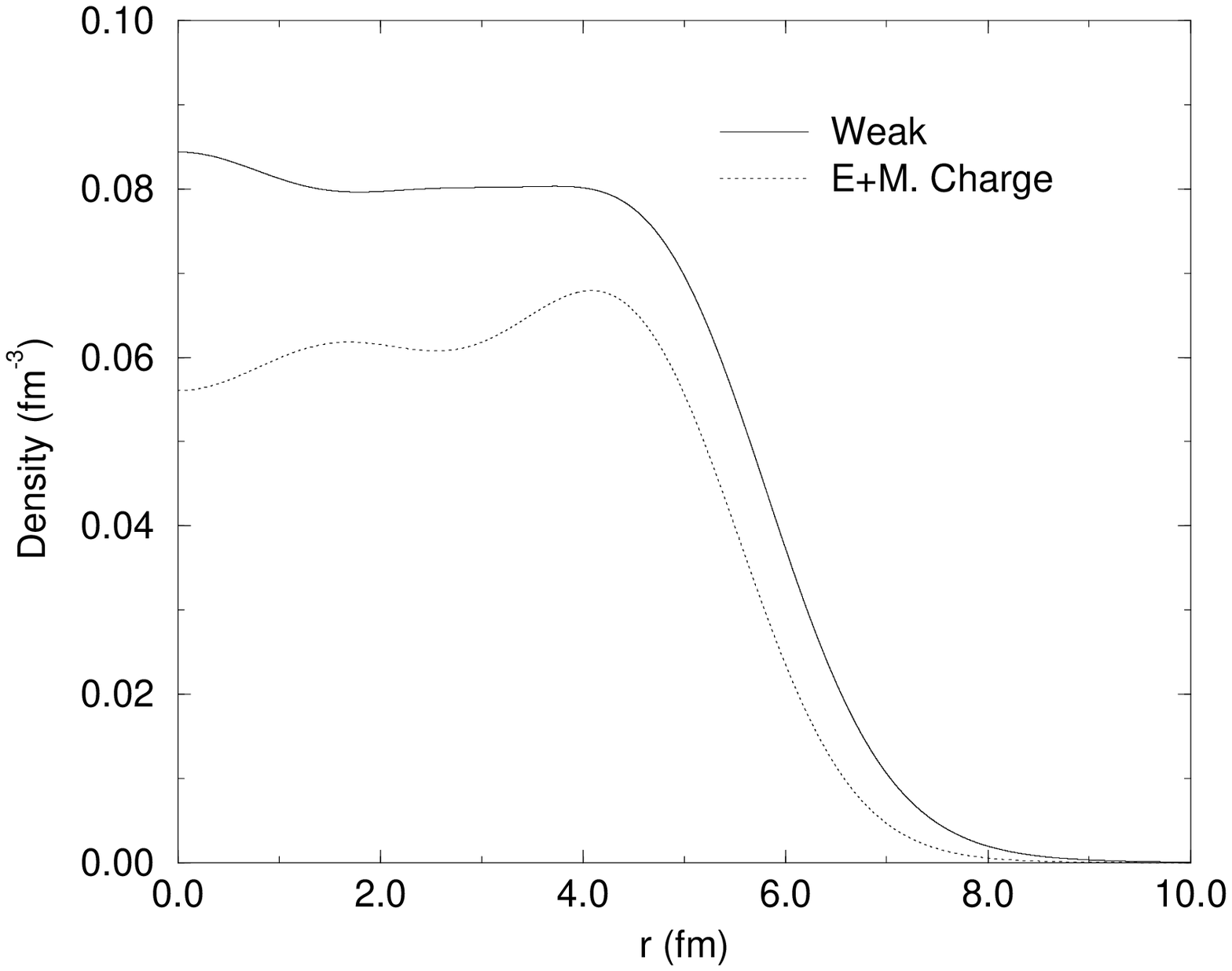,width=6in}}
\caption{Densities of $^{138}$Ba vs. radius $r$ for a relativistic mean field 
calculation [10].  The solid curve is minus the weak density while the charge
density is dotted.}\end{figure}

\begin{figure}
\centering{\ \psfig{figure=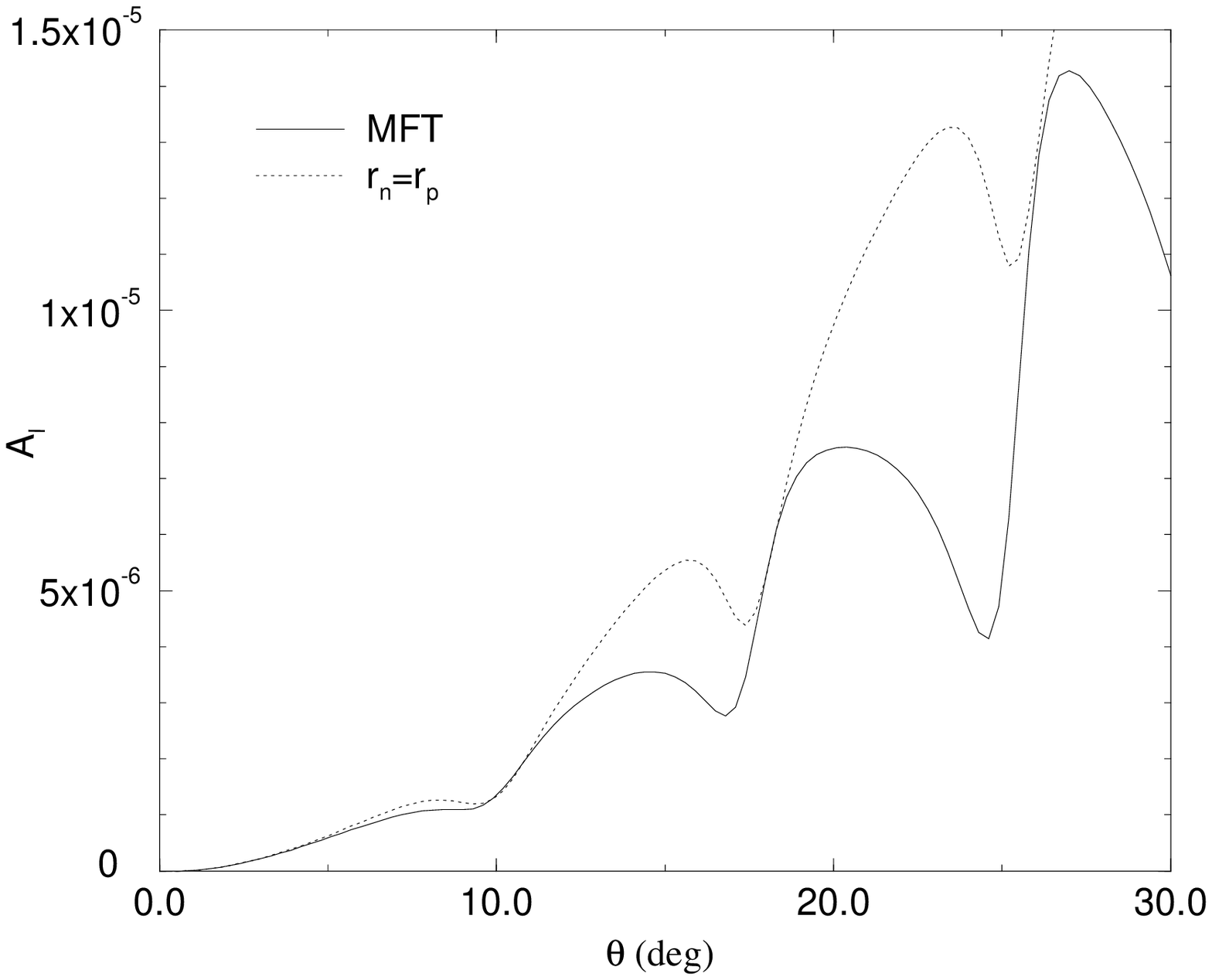,width=6in} }
\caption{Parity violating asymmetry $A_l$ for $^{138}$Ba 
at 850 MeV vs. scattering angle $\theta$.  
The solid curve is based on relativistic mean field densities
while the dotted curve assumes equal weak and charge densities.}\end{figure}

\begin{figure}
\centering{\ \psfig{figure=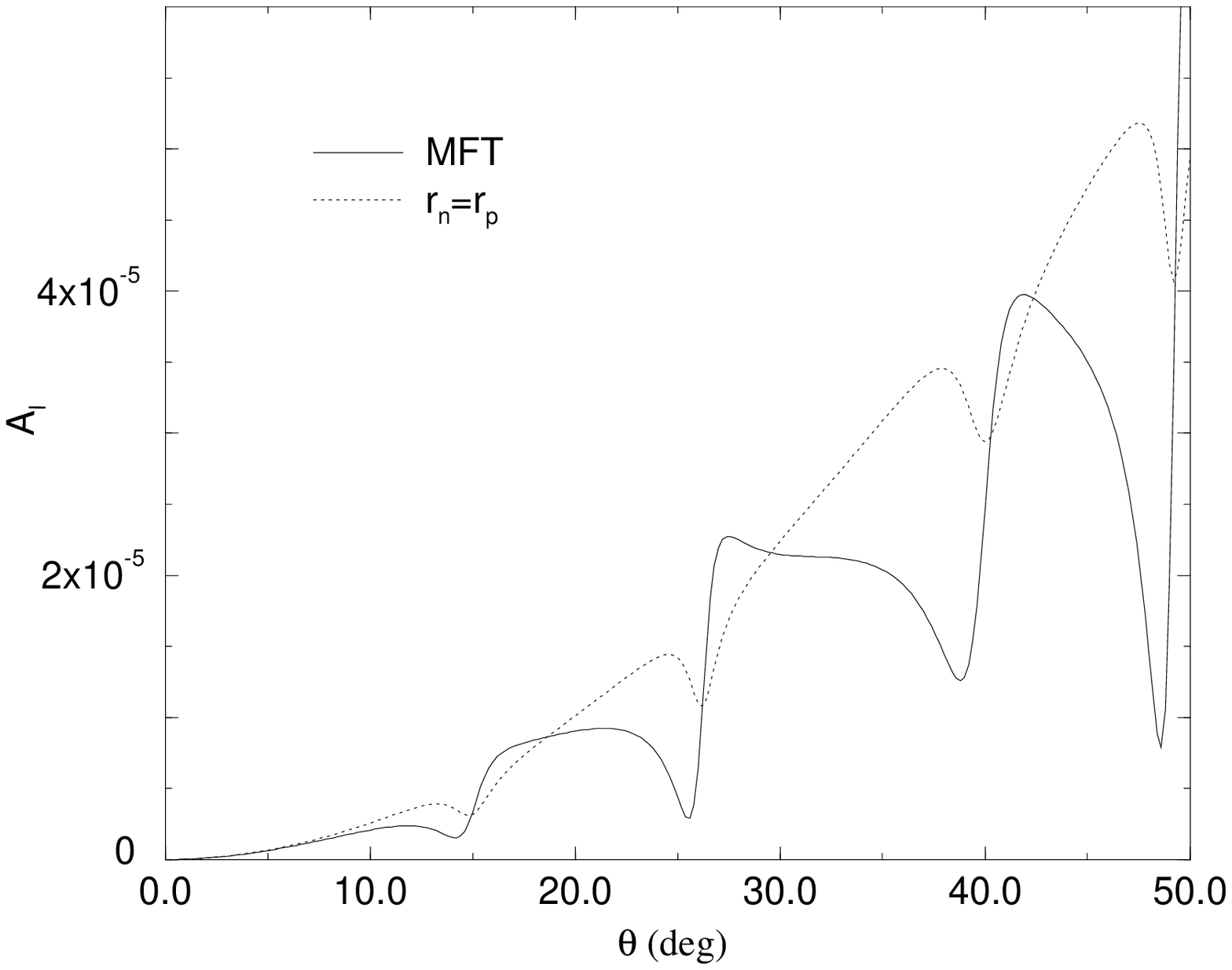,width=6in} }
\caption{Parity violating asymmetry $A_l$ for $^{48}$Ca at 850 MeV vs. 
scattering angle $\theta$.  
The solid curve is based on relativistic mean field densities
while the dotted curve assumes equal weak and charge densities.}
\end{figure}

\begin{figure}
\centering{\ \psfig{figure=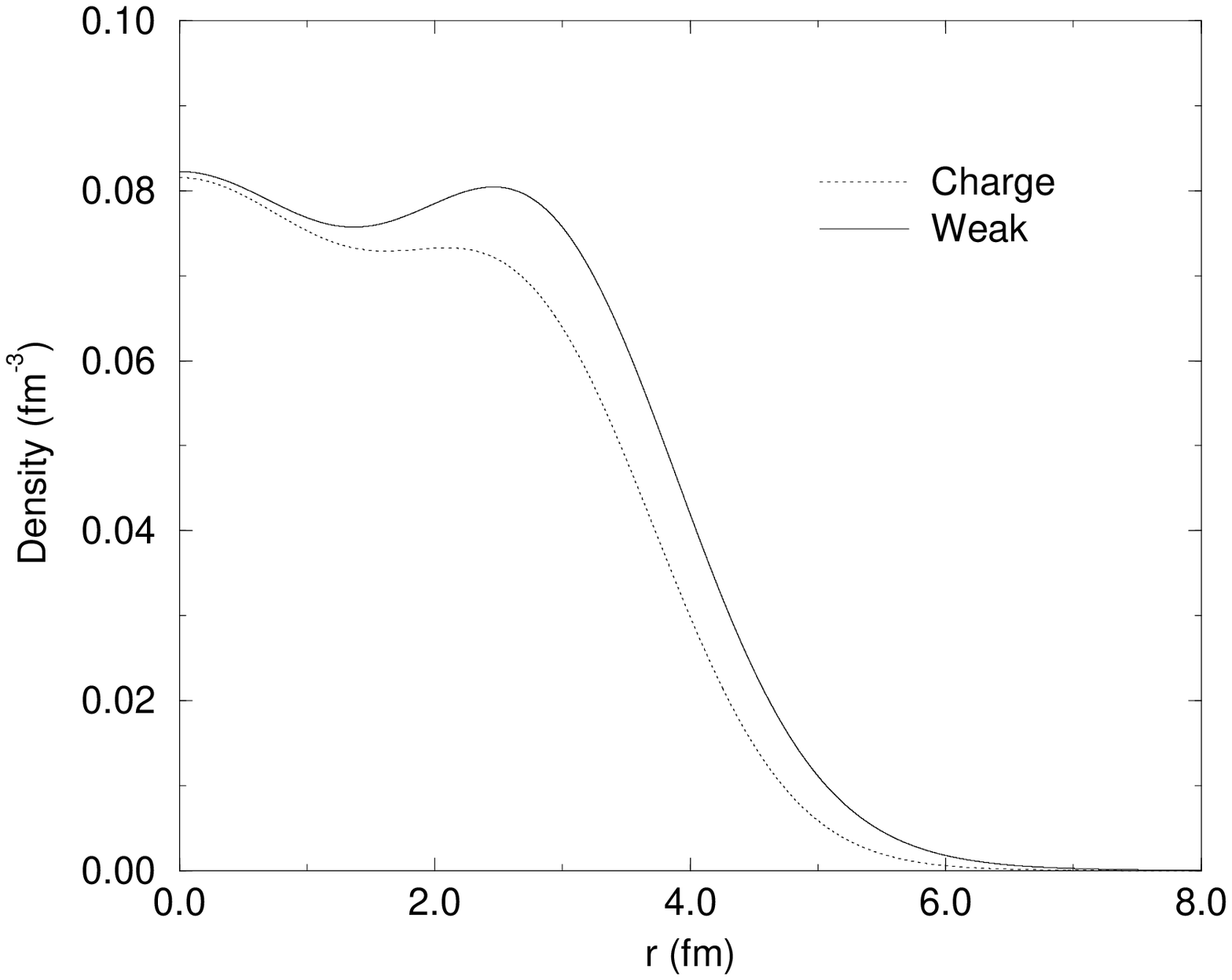,width=6in} }
\caption{Densities of $^{48}$Ca vs. radius $r$ for a relativistic mean field 
calculation [10].  The solid curve is minus the weak density while the charge
density is dotted.}
\end{figure}

\begin{figure}
\centering{\ \psfig{figure=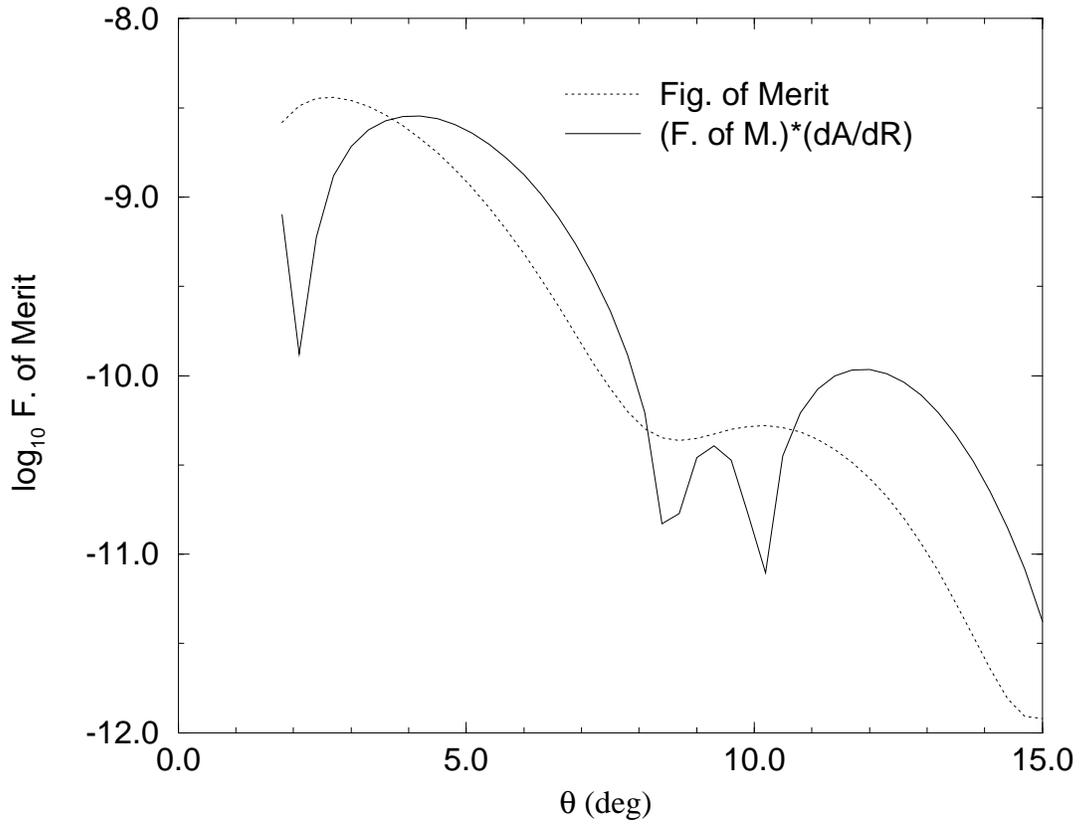,width=6in}}
\caption{Figure of merit, dotted curve, (differential cross section 
times asymmetry $A_l$ squared) log$_{10}$ in mb/Sr vs. scattering angle 
$\theta$ for $^{208}$Pb at 850 MeV.  The solid curve is the figure of merit 
multiplied by the logarithmic derivative of the asymmetry with respect
to the neutron radius (see Fig. 13).}\end{figure}

\begin{figure}
\centering{\ \psfig{figure=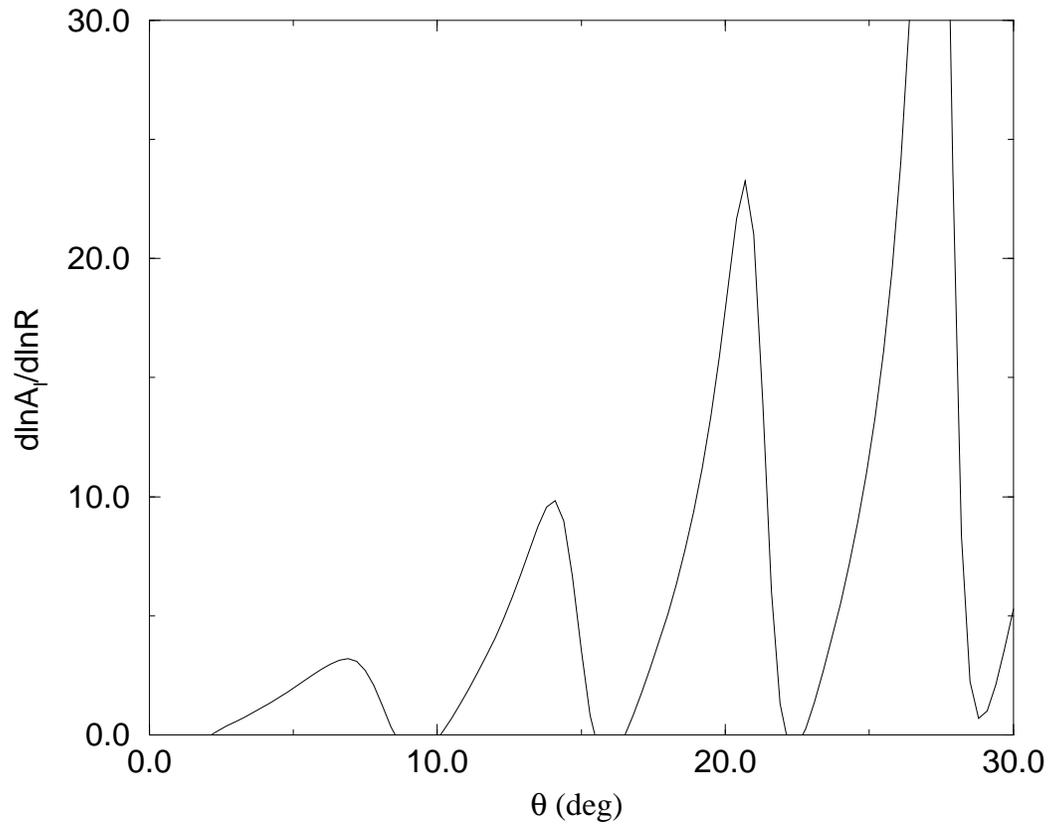,width=6in}}
\caption{Logarithmic derivative of the asymmetry $A_l$ with respect to
the neutron radius for $^{208}$Pb at 850 MeV vs. scattering angle $\theta$.
}\end{figure}
 
\begin{figure}
\centering{\ \psfig{figure=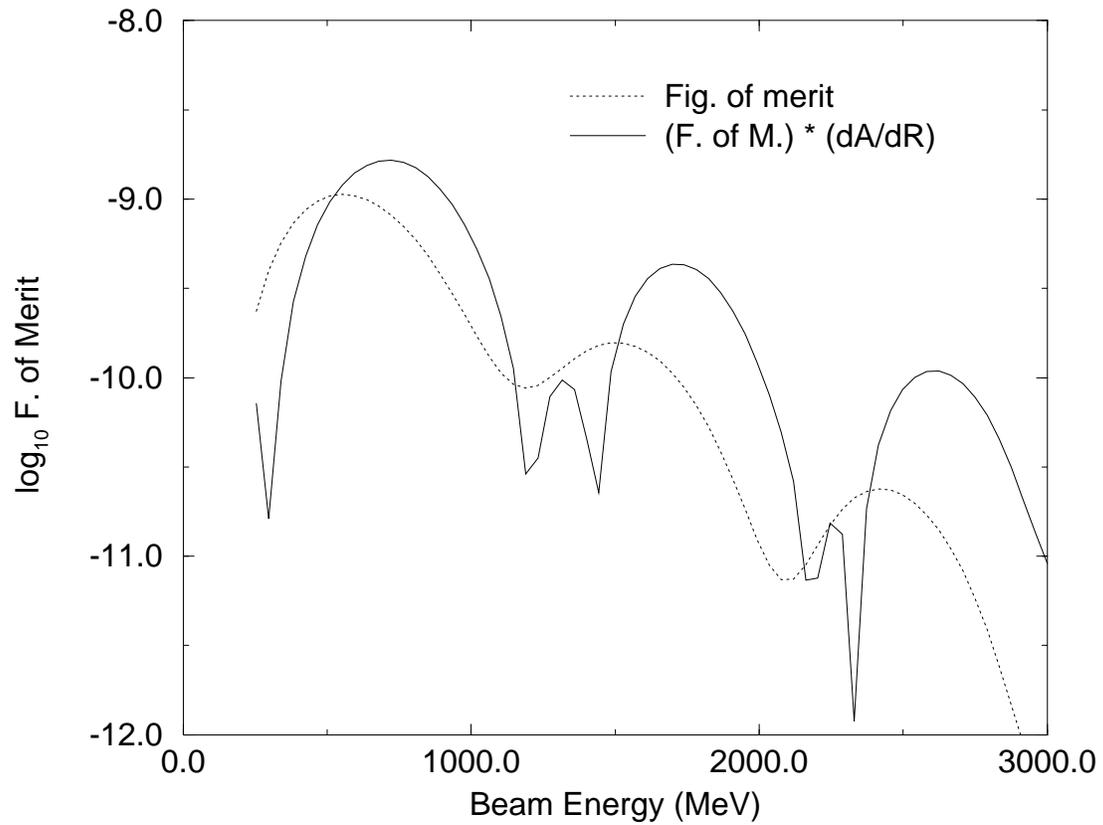,width=6in} }
\caption{Figure of merit, dotted curve, vs. beam energy  for a fixed 
laboratory scattering angle of six degrees for $^{208}$Pb.  The solid 
curve is the product of the figure of merit times the logarithmic 
derivative of the asymmetry with respect to the neutron radius.  
Note these curves are approximate.  They are based on distortions 
calculated at 850 MeV and assumed independent of energy.} \end{figure}

\end{document}